# Population genomics on the fly: recent advances in *Drosophila*

Annabelle Haudry[1]*, Stefan Laurent[2], Martin Kapun[3]

[1] Université de Lyon, Université Lyon 1, CNRS, Laboratoire de Biométrie et Biologie Evolutive UMR 5558, F-69622 Villeurbanne, France

[2] Department of Comparative Development and Genetics, Max Planck Institute for Plant Breeding Research, 50829 Cologne, Germany

[3] Department of Biology, University of Fribourg, CH-1700 Fribourg, Switzerland

* corresponding author: annabelle.haudry@univ-lyon1.fr

**Running head**

**Abstract**

*Drosophila melanogaster*, a small dipteran of African origin, represents one of the best-studied model organisms. Early work in this system has uniquely shed light on the basic principles of genetics and resulted in a versatile collection of genetic tools that allow to uncover mechanistic links between genotype and phenotype. Moreover, given its world-wide distribution in diverse habitats and its moderate genome-size, *Drosophila* has proven very powerful for population genetics inference and was one of the first eukaryotes whose genome was fully sequenced. In this book-chapter, we provide a brief historical overview of research in *Drosophila* and then focus on recent advances during the genomic era. After describing different types and sources of genomic data, we discuss mechanisms of neutral evolution including the demographic history of *Drosophila* and the effects of recombination and biased gene conversion. Then, we review recent advances in detecting genome-wide signals of selection, such as soft and hard selective sweeps. We further provide a brief introduction to background selection, selection of non-coding DNA and codon usage and focus on the role of structural variants, such as transposable elements and chromosomal inversions, during the adaptive process. Finally, we discuss how genomic data helps to dissect neutral and adaptive evolutionary mechanisms that shape genetic and phenotypic variation in natural populations along environmental gradients. In summary, this book chapter serves as a starting point to *Drosophila* population genomics and provides an introduction to the system and an overview to data sources, important population genetic concepts and recent advances in the field.

**Key Words**

# 1. Introduction

The fruit fly *Drosophila melanogaster* is a small Dipteran that originates from sub-Saharan Africa [1] and has since then colonized all continents except for Antarctica as a human commensal [2, 3]. Within the last 15-20,000 years it expanded its range to Europe and Asia and was only recently introduced to Australia and the Americas (~200 years ago according to [1, 4]). Because of its short life cycle and its simple maintenance, it was first adopted as a laboratory model organism by William Castle and later by Thomas Hunt Morgan at the beginning of the 20$^{th}$ century [3, 5]. At a time when the basic principles of heredity were still under heavy debate, Morgan used the *Drosophila* system to experimentally prove and extend the fundamental predictions of Mendelian genetics, which led to the discovery of genes and their location on chromosomes. This early work was rewarded with Nobel prizes to Morgan and several of his former students and research assistants and forms the basis of our present day understanding of genetic mechanisms [6]. Subsequently, the *Drosophila* system was further exploited, and resulted in the development of numerous genetic tools such as balancer chromosomes, gene-specific knock-out mutants and other transgenic constructs, including the Gal4/UAS system to study gene expression or more recently, the CRISPR/Cas9 system for site-specific genome engineering. Moreover, with its condensed genome of ~180Mb, *D. melanogaster* was among the first eukaryotic organisms whose genome was fully sequenced, assembled and annotated [7].

Beside major advances in functional genetics *Drosophila* has also proven powerful for population genetic inference. Accordingly, numerous major population genetics discoveries have first been made in flies. Theodosius Dobzhansky, together with co-workers and students, was one of the first to systematically investigate genetic variation in *Drosophila* - particularly by focusing on chromosomal inversions. His ground-breaking work gave a first insight into the evolutionary processes that shape genetic variation and subsequently paved the ground for the modern synthesis of evolutionary biology (Note1) [8, 9]. By sequencing the *Adh* gene in 11 lines collected in 5 natural populations, Hudson generated the first fruit fly DNA sequence polymorphism data, identifying only one non-synonymous polymorphism out of 43 SNPs [10]. As early as the 1980s, methods based on restriction enzymes were applied to *D. melanogaster* to quantify natural genetic variation across multiple loci [11, 12], followed by the first analyses of Sanger sequenced DNA fragments from dozens of genes [13]. These studies provided the first insights into genome-wide patterns of variation in DNA sequences, revealing abundant silent nucleotide site diversity, less abundant nonsynonymous diversity and rarer small insertions and deletions and transposable element insertions [14]. Based on the null hypothesis of neutral evolution, Hudson et al. proposed a first statistical test of selection based on comparing polymorphism and divergence: the Hudson–Kreitman–Aguadé (HKA) test [15], which postulates that genes should all exhibit the same ratio of within-species variability (polymorphism) to between-species divergence at neutral sites. As an extension of the HKA test, McDonald and Kreitman developed a novel test to specifically detect positive selection on protein sequences, first used to detect positive selection at the *Adh* locus in *Drosophila*, and which has since become a ubiquitous test of neutrality [16]. The ratio of non-synonymous to synonymous divergence is expected to be equal to the ratio of non-synonymous to synonymous polymorphism if non-synonymous sites are neutral or deleterious, but higher if they are adaptive. Some of the strongest evidence for adaptive molecular evolution documented in all organisms has come from application of the McDonald-Kreitman test and methods based on it (reviewed in [17, 18]). Finally, a major discovery made in *D. melanogaster* was that the level of nucleotide variability is positively correlated with the local recombination rate [19], suggesting that selection may constitute a major constraint on levels of genomic diversity.

In summary, the fruit fly *D. melanogaster* is an ideal model for studying neutral and adaptive genome evolution in outbred, sexual organisms since it is characterized by a long history as a genetic model organism [5], exhibits well-documented, rapid and widespread adaptations over short (<20 generations) timescales in natural populations [20, 21], has powerful genetic tools [5, 22] a well-annotated genome [23], and genome-wide polymorphisms data for several populations (see 2.2 for details). Moreover, the genomes of over 25 of its congeners have been recently sequenced [24]. In particular, comparative genomics analyses on 12 species provided fundamental new insights into genome evolution [25] and led to the ModENCODE Project [26], which aims at identifying functional elements in the *D. melanogaster* and *Caenorhabditis elegans* genomes. In this chapter, we will focus on population genomics studies (Note 2), mostly based on next generation sequencing data, and review different aspects of both neutral and selective evolution based on the *Drosophila* system.

## 2. Data sources

### 2.1. Data acquisition techniques

One particular strength of the *Drosophila* system is its simple maintenance under laboratory conditions. *Drosophila* is commonly propagated as isofemale lines which originate from a single wild-caught and inseminated female. This allows researchers to conduct molecular and phenotypic measurements across several years using the same genetic material and to preserve natural genetic variation under laboratory conditions. In this paragraph we briefly review the nature of the genetic material that has been sequenced in large genome sequencing projects and how these different approaches potentially affect patterns of variation and missing data.

### 2.1.1. Isofemale inbred lines

Isofemale inbred lines are started from single gravid females whose progeny are allowed to interbreed. These lines can be maintained for several years as long as flies are regularly transferred to new vials with fresh fly-food (a well-known task for any student in a *Drosophila* lab having worked in a fly-room). A high degree of inbreeding due to small population sizes leads to a rapid reduction of genetic variation and heterozygosity at every generation within each isofemale line. Inbred lines are often referred to as *F* (Filial generations) followed by the number of the generations of full-sib mating (*F3*, *F10*, *F20*...). Due to their near-complete homozygosity, every line should be considered as contributing a single genome to the total sample (and not two, as it could be assumed for an outbred sample). Since isofemale lines are propagated separately and are not allowed to interbreed, they are a versatile tool to preserve genetic variation under laboratory conditions, given that sufficient isofemale lines per population are maintained [27]. One significant issue with this approach is that lines derived from equatorial populations have shown to be particularly resistant to inbreeding, a problem that has been linked to the presence of inversion polymorphisms hosting recessive lethal mutations. In these lines, large regions (>500 kb) of residual heterozygosity can be observed [28] which complicates the determination of patterns of polymorphism and divergence in this population [29, 30]. Moreover, given the small population sizes at which isofemale lines are usually propagated, novel mutations that appeared after the capture of the wild-caught ancestors are likely to accumulate in each line over time. Isofemale lines that are maintained in the laboratory for long periods of time will thus slightly deviate from their ancestors and be poorer indicators of natural variation compared to recently established lines.

### 2.1.2. Haploid embryo sequencing

To circumvent problems caused by residual heterozygosity, Langley et al. proposed to sequence the amplified genome of a single haploid embryo [29]. Most eggs fertilised by recessive male sterile mutants *ms(3)K81* fail to develop [31]. The few that do, however, only contain one haploid maternal genome. Such a single haploid embryo derived from a cross between a female from any line of interest and a *ms(3)K81* male provides enough genomic DNA for whole-genome amplification and sequencing [29]. Although whole-genome amplification increases variance in coverage and the frequency of chimeric reads, this technique provides a powerful approach to uniquely generate high-quality sequencing data using standard paired-end sequencing protocols. Similar to isofemale inbred lines, this technique provides a single genome per sequenced individual (female) but allows obtaining phased DNA sequences even in the presence of inbreeding-resistant polymorphic inversions.

### 2.1.3. Genomic sequencing and phasing of hemiclones

Whole genome sequencing of hybrid F1 crosses – so called hemiclones – which share one common parent [32], represents an alternative approach to generate phased haplotype sequencing data. Wild-type *Drosophila* strains are therefore crossed with the same highly inbred or fully isogenic lab-strain that acts as a reference. The resulting F1 hemiclones are then sequenced as single individuals alongside their lab-strain parent to bioinformatically distinguish between the reference and the unknown wild-type allele. This method has been recently employed in *D. melanogaster* and allowed to combine cytological screens with whole genome sequencing to generate and analyze fully phased genomes with known inversion polymorphisms [33]. Additionally, this approach was used to sequence and characterize a panel of more than 200 wild-type chromosomes from a North American *D. melanogaster* population [34].

### 2.1.4. Pooled sequencing (Pool-Seq)

Pool-Seq is a sequencing technique, where tissues or whole bodies of multiple individuals are pooled prior to DNA extraction, library preparation and whole genome sequencing. In contrast to single individual sequencing, Pool-Seq is very cost-efficient and has proven powerful to accurately estimate population-wide allele frequencies [35–37]. However, Pool-Seq also comes at the cost of losing information about individual genotypes and haplotype structure. Moreover, it remains very difficult to distinguish low-frequency variants from sequencing errors, which further complicates population genetics inference [38, 39] and precludes calculating classic population genetic estimators without statistical adjustments (see for example [40–43]).

It is important to note that these approaches neither allow to measure genotype variation in natural populations, which is the proportion of heterozygote individuals within a population nor the proportion of heterozygote sites within a single diploid individual.

## 2.2. Consortia and available datasets

The first finished genome draft of *D. melanogaster* was published more than 17 years ago, and was among the very first fully sequenced eukaryotic genomes [7]. Since then, the quality of the reference sequence has further improved, and the number of functional annotations, such as gene models or regulatory elements, keeps increasing continuously. Both sequence and annotation data are publicly available at www.flybase.org, a bioinformatics database that is the main repository of genetic and molecular information for *D. melanogaster* (and other species from the Drosophilidae family). *D. melanogaster* was also one of the first species for which full-genome intraspecific variation data was collected. The first whole-genome population genetics study in *D. melanogaster* surveyed natural variation in three African (Malawi) and six North American (North Carolina) strains using low-coverage sequencing [44].

### 2.2.1. *Drosophila* Genetic Reference Panel (DGRP) and *Drosophila* Population Genomics Project (DPGP)

The first two projects to systematically investigate the genomic variability in natural *D. melanogaster* populations were the DGRP [45] and DPGP [46] initiatives. Both consortia independently sequenced more than 160 isofemale inbred lines (F20), all sampled in Raleigh, North Carolina, USA; a sample that was later extended to 205 lines [47]. The major aim was to generate whole-genome sequencing data that can be used for genome-wide association studies. The genetic and phenotypic data are available from http://dgrp2.gnets.ncsu.edu. While the DGRP data are well suited for quantitative genetics analyses (using stable, well-described, and homogeneous genetic material), they only provide information about the genetic variation at a single location (North-Eastern US) although a large portion of the genetic diversity of the species is known to reside in its ancestral range in sub-Sahara Africa [48, 49]. The DGRP data is thus neither suitable for investigating the demographic history of worldwide populations nor the patterns and processes leading to local adaptations that likely facilitated the range expansion and ultimately led to a cosmopolitan distribution of *D. melanogaster*.

### 2.2.2. Drosophila Population Genomics Projects

The *Drosophila* population genomic project (DPGP, http://www.dpgp.org) is an ongoing major population genomic sequencing effort: beside the Raleigh population, the DPGP sequenced a population of Malawi (Africa) that exhibited > 40% more polymorphism genome-wide compared to the North-American one [46]. Then, the DPGP2 sequenced 139 wild-derivates strains representing 22 populations from sub-Saharan Africa [50]. The analyses of the DPGP2 data confirmed that the most genetically diverse populations are located in Southern Africa (e.g. Zambia). Afterwards, the DPGP3 increased the sample size for a Zambian population (Siavonga) up to 197 lines [51]. Most DPGP2 and all DPGP3 lines were sequenced from haploid embryos as described above.

### 2.2.3. The Drosophila Genome Nexus

The Drosophila Genome Nexus is a population genomic resource that integrates single-individual *D. melanogaster* genomes from multiple published sources [51, 52], including DPGP and DGRP amongst others [30, 53–55]. The aim was to generate a comprehensive dataset using the same bioinformatics methods to facilitate comparisons among them. The latest iteration (DGN v.1.1 [52]), contains a total of 1,121 genomes, from 83 populations in Africa, Europe, North America and Australia. It especially highlighted differences in levels of heterozygosity among the different datasets. The genome browser PopFly allows the visualization and retrieval of numerous population genomics statistics, such as estimates of nucleotide diversity, linkage disequilibrium, recombination rates [56].

### 2.2.4. Dros-RTEC and DrosEU

Complementary to previous efforts, which aim at sequencing single individual genomes in large numbers from a single population (DGRP, DPGP, DPGP3) or in small numbers from multiple locations (DPGP2 [30]), two consortia in North America (Dros-RTEC [57]) and in Europe (DrosEU [43]) recently started to generate Pool-Seq data from wild-caught flies from numerous sampling sites to quantitatively assess genetic variation and differentiation through time and space in natural populations. To date, DrosEU has sequenced and analyzed 48 samples from more than 30 localities all across Europe, which revealed strong and previously unknown population structure - mostly along the longitudinal axis - in Europe. Moreover, population genetic analyses of these data allowed a description of novel candidates for selective sweeps, to detect previously unknown clines of mitochondrial haplotypes, inversions and transposable elements (TE) and to isolate novel viral species in the microbiome. The Dros-RTEC consortium similarly sequenced 72 samples of *D. melanogaster* collected from 23 localities mostly in North America [57]. Due to their focus on rapid seasonal adaptation, many localities were sampled at different timepoints over the course of one to six years, which allows a quantitative investigation of genome-wide seasonal fluctuations in SNPs and inversion polymorphisms. These analyses revealed that previous candidates for seasonality exhibit highly predictable annual allele frequency fluctuations and those signatures of seasonal adaptation parallel spatial differentiation along latitudinal gradients.

### 2.2.5. Other data

In addition to these concerted sampling and sequencing efforts, there is a rapidly growing number of studies that similarly sequenced pools of flies from natural populations. For example, Pool-Seq data of populations from the temperature gradients along the North American and Australia were generated [58–60]. Large pools of flies collected from Vienna/Austria and Bolzano/Italy were sequenced by [61]. More recently, Kofler and colleagues [62] generated and analyzed Pool-Seq data from more than 550 South African flies. In combination with the aforementioned Pool-Seq data from large consortia, these data represent highly valuable resources to tackle fundamental questions about the adaptive process on complex spatial and temporal scales.

## 3. Neutral evolution

### 3.1. Demographic analyses

*D. melanogaster* is one of eight species described in the melanogaster subgroup of the subgenus *Sophophora*. Within this group, two species are cosmopolitan (*D. melanogaster* and *D. simulans*), while the remaining six are endemic to the Afrotropical region (*D. sechellia*, *D. mauritania*, *D. erecta*, *D. orena*, *D. teissieri*, *D. yakuba*). This has led early studies to suggest an Afrotropical origin of *D. melanogaster* and *D. simulans* and is now widely accepted [4]. As expected under this hypothesis, the genome-wide average diversity measured in Afrotropical populations of *D. melanogaster* is higher than in non-African populations [44, 48, 63, 64]. In addition and similarly to *Homo sapiens*, the genetic variation outside sub-Saharan Africa represents a subset of the diversity found within sub-Saharan African populations, which further suggest that South-Eastern tropical Africa represents the ancestral range of the species [65].

In an influential review summarizing the results of early population variation surveys in *D. melanogaster* [4], David and Capy categorized world-wide natural populations into three groups: ancestral, ancient, and new populations (Figure 1). Ancestral populations are located in sub-Saharan Africa, where they probably have split from the sister species *D. simulans* approximatively 2.3 million years ago [66]. Ancient populations are located in Eurasia and migrated out of their ancestral range presumably at the end of the last ice age. The third group, new populations, are located in America and Australia, and represent a blend of ancestral and ancient populations that recently colonized these two continents along European shipping routes during the last centuries.

Although these early insights were based on a small number of loci, they have proven to be surprisingly robust and 30 years later, the categorization of David and Capy is still widely accepted. Several studies, however, took advantage of the increasing amount of genetic data and the rapidly developing field of model-based inference in population genetics, to investigate the demographic history of the species within a probabilistic framework. These studies evaluated the likelihood of competing demographic scenarios and provided estimates for demographic parameters such as the age of the split between African and non-African populations, and population sizes at different times of the colonization process. In the next paragraph we review how genome-wide data and statistical modeling updated the insights formulated by David and Capy [4].

Early population genetics surveys identified East and South African populations to be closer to mutation-drift equilibrium compared to West African populations, which were characterized by higher linkage disequilibrium and lower diversity levels [63–65]. These findings suggest that East and South Africa include the ancestral range of the species. Demographic inference using samples from sub-Saharan Africa indicated that the ancestral population has experienced a population size expansion approximately 60,000 years ago (ranging from 26,000 to 95,000 [67]). This ancestral expansion is found in all published models incorporating the African population and is necessary to fit the excess of rare variants measured in samples from the ancestral range (e.g. Zambia, Zimbabwe). These models, however, assume that all sampled mutations are neutral, which is unlikely because of putatively unknown regulatory elements and the presence of background selection [68]. A simulation showed that ignoring background selection in demographic inference leads to an overestimation of growth models [69]. Estimations of the coalescence rate through time using smc++ indicated that the rate of coalescence in a sample from the Zambian population (Siavonga, DPGP3) has been constantly decreasing in the last 100,000 years [70], which is in line with the population expansion scenario suggested by previous studies. Furthermore, Terhorst et al. [70] measured a strong reduction in the coalescent rate for times older than 100,000 years, suggesting either a very large ancestral population size or substantial population structure in the ancestral population [71]. Neither of these two processes is accounted for in current demographic models for *D. melanogaster* and more work is needed to evaluate whether the decreased ancestral rate measured by [70] is reflecting true ancestral processes or rather aspects of the genomic data that are not accounted by the method. More specifically, it remains to be clarified whether this approach can correctly recover neutral demographic processes when applied to small compact genomes with a high proportion of non-neutral regions. More recently, Kapopoulou et al. [72] estimated the age of the split between ancestral (Zambia) and West African populations to be approximately 72,000 years, which suggests that the population expansion reported by earlier studies could well reflect a genuine early range expansion of the species on the African continent.

### 3.1.1. *Out of Africa*

Analysis of European samples revealed that the time of split between African and European populations occurred around 13,000 years [67, 73]. These early studies, however, did not include gene flow between populations in their models and therefore predicted that their estimates were probably younger than the true age of divergence between African and European lineages. Indeed, Kapopoulou et al. [74] recently confirmed this prediction using genome-wide polymorphism data and by explicitly accounting for the effect of gene flow in their inference procedure. Their demographic results identified gene flow as an important factor in the recent history of European and African populations and reported divergence time estimates of approximately 48,000 years. Independently, Pool et al. [50] reported pervasive influence of European admixture in many African populations with greater admixture proportion in urban locations. The "ancient" status of Southeast Asian populations has also been confirmed by [73] and Arguello et al (2018). Similarly to the European case described above, divergence time estimates between Asian and European populations strongly depend on whether or not gene flow is taken into account in the inference method (22,000 vs. 5,000 years, respectively).

North-American populations are considered as newly introduced because the colonization process has been observed directly by entomologists in the second half of the 19[th] century [2]. Strikingly, *D. melanogaster* was identified as the most common species across the United States only 25 years after its introduction, suggesting a dramatic population expansion after colonization [2]. A genome-wide analysis of 39 flies sampled as part of the DGRP project [45], using an approximate Bayesian computation method (ABC) revealed the admixed nature of this population with European and African admixture proportions of 85 and 15%, respectively [75]. These estimations confirmed similar conclusions reached earlier using microsatellite data [76]. This very recent secondary contact between African and European lineages is likely responsible for the North-South clinal genetic variation observed in Northern-America and Australia (Figure 1), but local adaptation could contribute to the maintenance of this clinal variation by opposing itself to the homogenizing effect of gene flow [54, 77]. Arguello et al. (2018) recently confirmed the importance of Afro-European admixture in the ancestry of North American and Australian flies using a larger dataset and a more precise inference procedure. The mosaic ancestry of American and Australian fly populations therefore represents an exciting opportunity to study how migration and selection interact along a clinal heterogeneous environment. Methods based on hidden Markov models were developed to estimate patterns of local ancestry in samples of North-American populations (where the term local refers to an arbitrary sub-genomic unit) [78, 79]. In samples with a predominant European genetic background, their results identify significant differences in the proportion of African ancestry between functional classes of genomic loci.

**3.2. Recombination**

In most sexually reproducing eukaryotes, recombination ensures both the proper segregation of homologous chromosomes during meiosis and the creation of new combinations of alleles at each generation. During meiosis, a substantial number of double-strand breaks result in meiotic recombination between homologs. These double-strand breaks are repaired either as crossover (CO) or non-crossover (NCO) gene conversions: COs imply reciprocal exchange between flanking regions, whereas NCOs do not. Both forms of recombination are key factors in genome evolution as their rates determine the probability to which extent genomic sites are linked or evolve independently and hence affect the evolutionary fate of the alleles. A fundamental understanding of recombination rates is thus crucial in population genomic studies. In *Drosophila*, meiotic recombination only occurs in females, but not in males [80], a dimorphism known as "achiasmy" (an extreme case of heterochiasmy observed in many species [81]).

In the 1990's, several studies revolutionized population genetics by showing that the level of genetic diversity in populations of *Drosophila* species was lower in regions of low recombination [19, 82, 83]. Recombination itself seems to be the major factor determining patterns of nucleotide diversity along the genome. Indeed, mutation associated with recombination can be excluded as the cause of this correlation, at least in *Drosophila*, given the lack of correlation between recombination and divergence [19, 45]. The frequent occurrence of these patterns [84] has motivated further exploration and estimation of genome-wide patterns of recombination and diversity.

Classically, the estimation of recombination rates generally relies on the "Marey approach" that compares a genetic map, which quantifies distances as CO frequency (in cM) to a physical map (distances in basepairs). A user-friendly web service called MareyMap Online [85] allows to get recombination rate estimates based on such an approach. In their landmark study, Begun and Aquadro [19] found a strong positive correlation between nucleotide diversity estimated at 20 genes and local rates of COs in natural populations of *D. melanogaster*. They used the coefficient of exchange as a measure of recombination rate, based on the physical distance among cytological markers in combination with DNA content estimates from densities of polytene chromosomes [86]. The fully sequenced *Drosophila* genome, which became available in 2000 [7], represents a highly accurate physical map that was necessary to generate detailed recombination maps. Marais, Mouchiroud and Duret [87] fitted a third-order polynomial, which provided a first overview of the distribution of COs along each chromosomal arm. They showed that CO rates decline in proximity to telomeres and centromeres. Accounting for specific recombination patterns of the telomeric and centromeric regions, Fiston-Lavier and colleagues provided corrected estimates of local recombination rates in *D. melanogaster* [88].

Besides classical recombination maps based on crosses, alternative approaches take population genetic variation into account to estimate CO rates. Patterns of linkage disequilibrium (LD) in a population result from historical recombination events. Recombination (CO) rates across the genome can thus be inferred from linkage disequilibrium, through the population-scaled recombination parameter $\rho = 4N_e r$ where $N_e$ is the effective population size, and $r$ the CO rate between base pairs per generation [89]. Mc Vean et al. [90] developed a coalescent-based method implemented in the software *LDHat* for the estimation of local recombination rates ($4N_e r$ per kilobase) using a composite likelihood approximation [91], based on the segregation of a high density of physically mapped SNPs. Originally developed for human populations, this method has been applied to many species including *Drosophila* [46]. Besides providing a recombination map with a higher resolution, Langley et al. showed that $r$ and $\rho$ were strongly positively correlated at a large scale [46], indicating these independent estimates are both capturing heterogeneity in recombination. However, compared to humans, *D. melanogaster* harbours much higher SNP densities, population recombination parameters are an order of magnitude higher and footprints of positive selection are more widespread. Since the *LDhat* method assumes neutral evolution, it can infer spurious recombination hotspots under certain conditions of selection. Chan et al. [92] proposed a corrected method (*LDhelmet*), which is more robust to the effects of selection and computed an improved fine-scale, genome wide recombination map in *D. melanogaster*, including a handful of hotspots of at least ten times the background recombination rate.

Combining both crosses and population variation approaches, Comeron et al. [93] proposed a method to distinguish between the two possible outcomes of the repairing of double strand breaks associated with meiotic recombination: CO and NCO gene conversions. While COs involve DNA exchange between chromatid arms of homologous chromosomes on a large-scale, NCOs are non-reciprocal recombination events with a swap of small DNA fragment. First described in *Drosophila* [94–96], CO interference prevents the formation of two COs in close proximity and thus, reduces the probability of double CO events (~1 CO per chromosome per meiosis [97]). Based on the size of genetic regions affected by gene conversion, Comeron et al. [93] estimated separately rates of CO and NCO from crosses, making use of the very high density of SNPs in *D. melanogaster* (139 million), which allowed them to design a 2kb-resolution map of recombination. Unlike COs, NCOs appear to be uniformly distributed throughout the genome [93], insensitive to the centromere effect and without interference [98], and more frequent (rates of NCO: CO could reach values over 100 [93]).

While extrapolated and direct recombination estimates are consistent on a large scale, the latter ones show greater variability at the center of the chromosomal arms [99]. Altogether, these recombination maps provide baseline estimates for population genomic studies, especially to model the expected variation under selection at linked sites (see paragraph 4.2 and [100]).

**3.3. Biased Gene Conversion**

Both CO and NCO recombination involve gene conversion. In particular, the presence of heterozygous sites within heteroduplex DNA results in the formation of mismatches, which lead to the conversion of one allele by the other during the repair. There is evidence, from diverse eukaryotic lineages, that GC:AT mismatches tend to be more often repaired in GC than in AT alleles, a process called GC-biased gene conversion (gBGC [101, 102]). gBGC has been inferred as the main driver of GC-content evolution in vertebrates [101, 103–105] and several other taxa [106–109]. gBGC is a non-adaptive mechanism that mimics natural selection, because it confers a higher transmission probability of GC over AT alleles in heterozygotes. Therefore, gBGC needs to be accounted for in molecular evolution studies to correctly model neutral evolution of the genome [110, 111]. The impact of gBGC in *D. melanogaster* is, however, less clear: GC content is positively correlated with CO rate [87, 112, 113], but not with NCO rate [93]. Globally, whole-genome polymorphism and divergence data did not support a gBGC model in *D. melanogaster* [114], except for the *X* chromosome [115, 116] where it may partly explain the stronger signal of selection on codon usage compared to autosomes [117].

### 3.4. Population genetics of chromosomal inversions

Chromosomal inversions were first discovered in *D. melanogaster* almost exactly 100 years ago [118]. They represent structural mutations that result in the reversal of genetic order in the affected genomic region relative to the non-inverted ('standard') arrangement [119, 120]. Inversions can have strong effects on genome evolution in various different ways: breakpoints may disrupt genes (e.g. [117]) or result in gene duplications due to staggered breaks [122, 123]. Moreover, inversions can trigger positional effects, where expression patterns of genes are altered due to changes in their relative chromosomal position ([124–126] but see [127]). However, their most fundamental effect is the strong suppression of recombination in heterozygotes, since crossing-over within the inverted region results in abnormal chromatids [128–130]. As shown in humans where inversions can cause numerous diseases, many of these effects have deleterious consequences [131]; however there are some rare adaptive cases (reviewed in 4.5). Upon their discovery, inversions have been predominantly studied in species of the genus *Drosophila*. Particularly the pioneering work of Theodosius Dobzhansky and colleagues in *D. pseudoobscura* and *D. persimilis* [8, 9, 132, 133] gave a first insight into the evolutionary processes that shape genetic variation and differentiation in natural populations [134–136]. However, only due to recent advances in whole-genome sequencing technology, it became possible to quantitatively test for different evolutionary models and characterize the genetic effects of inversions on a genome-wide scale. Consistent with the action of spatially varying selection, many inversions in *Drosophila* are commonly found to exhibit steep clines along environmental gradients [137–140]. Several of these, such as the latitudinal gradient of the well-studied *In(3R)Payne* inversion in *D. melanogaster*, are replicated on multiple continents and persist over time ([33, 141] but see [142]). Recent large-scale genomic datasets of *D. melanogaster,* for the first time, allow a quantitative assessment of the genetic and evolutionary pattern associated with inversions. Analyses of genome-wide data from African flies allowed for (1) determination of the age and geographic origin of various cosmopolitan and endemic inversions. These analyses revealed that most common cosmopolitan inversions are of African origin and predate the out-of-Africa migration [143]. Furthermore, (2) these data provide a first insight into the amount and distribution of genetic variation and differentiation associated with inversions. Data analyses of the DGRP, for example, found that inversions contribute strongest to genetic differentiation and substructure within a population from Raleigh/North Carolina [47]. Moreover, only with the help of dense genome-wide sequencing, it became possible to show that genetic differentiation is not homogeneously elevated within inversions, but decays towards the inversion center [33, 140, 143–145]. Consistent with theoretical predictions [146–148], these data suggest that there is a limited amount of genetic exchange among karyotypes rather than a complete inhibition of recombination. In addition, local peaks of strong differentiation close to the inversion center suggest that several inversions, for example *In(3R)Payne,* contain various adaptive loci which are in strong linkage with the inversion breakpoints [33, 140, 143, 144]. Analyses of genomic data in combination with long-range PCR further helped (3) to reconstruct the exact genetic composition of inversion breakpoints [149] and (4) facilitated the development of inversion-specific marker SNPs, which now make it possible to reliably estimate inversion abundance and frequency in single-individual and Pool-Seq data, respectively [33, 140, 150]. Together, these analyses highlight that whole-genome data for the first time allows to quantitatively elucidate the mechanisms underlying the evolution of chromosomal inversions.

**3.5. Population genomics of transposable elements**

Transposable elements (TEs) are mobile, self-replicating, repeated DNA sequences found in every eukaryotic genome at varying proportions among taxa [151, 152], among closely related species [24] and among individuals of the same species ([153] for maize; [154] for *Arabidopsis*; [155] for *Drosophila*). Because of their mutagenic potential (either by inserting into functional regions or by promoting chromosomal rearrangements via ectopic recombination -Note 3), TEs are thought to play a significant role in populations' evolution and adaptation [156]. According to the nearly neutral theory, TE insertions are expected to be generally neutral or deleterious to the host genome [157]. However, rare cases of adaptive insertions have also been documented (see 4.6 for examples in *Drosophila*). The general model of TE dynamics is the *transposition-selection balance model* [158]. It assumes that the maintenance of TEs in the population is explained by an equilibrium between (i) the increase in copy number through a constant transposition rate and (ii) their removal driven by natural selection, through the combined effect of excision and purifying selection acting against the deleterious effects of inserted TEs [158, 159]. This model predicts that most TEs should be segregating at low TE frequency in *D. melanogaster* populations (see [160] for detailed review). The *burst-transposition model* [161] relaxes the assumption of constant transposition rate over time in proposing periods of intense TE transposition (bursts) to explain TE dynamics. According to this model, recent insertions haven't yet reached an equilibrium between their transposition rate and negative selection. TEs may thus be at low frequency even under a strictly neutral model. Here, a positive correlation between insertions age and their frequency is expected (recently active TEs should be at low population frequency while long-time inactive TEs could reach fixation).

*D. melanogaster* has been used as a model species for the study of TE population dynamics for more than 25 years [162] and recent whole genome population data fuelled this area of research allowing testing of previous hypotheses. A bulk of new programs was recently developed to estimate TE insertion frequency in a population using NGS datasets (see [163] for review). On top of the 5,434 annotated TEs described in the reference genome, 10,208 and 17,639 insertions were discovered in European [164] and North American DGRP [165] populations, respectively. However, these numbers needs to be considered cautiously as the performance of methods detecting polymorphic TE insertions based on short read data depends on many variables, such as the sequencing coverage, the element family, the age of insertion, the size of the copy, the genomic location (see a benchmark in [166]). The large predominance of low frequency insertions along with the scarcity of insertions in exonic regions observed in both datasets support the *transposition-selection balance model*. In contrast, Kofler, Betancourt and Schlötterer provided evidence that half of the TE families have had transposition rates that vary with time [164], giving support to the *burst-transposition model*. However, they also found an excess of rare variants in young TE insertions compared to neutral expectations which suggests the action of purifying selection [165]. Overall, population genomics analyses of TEs provide empirical support for both hypotheses and indicate that they are not mutually exclusive. This is in agreement with previous *in situ* analyses suggesting that models of evolution could vary among elements and populations [167]. Although dynamics of some TE families can be explained by a neutral model with transposition rates varying over time, purifying selection is necessary to fully explain the patterns of population distribution of TEs [160, 168].

## 4. Selection

*D. melanogaster* has been a model species for many studies aiming at describing the genetic basis of adaptation. Comparisons between theoretical models of positive and negative selection with empirical data have started in the early 1980s, when PCR coupled with Sanger sequencing allowed to directly measure natural variation. The positive correlation between local rates of recombination and genetic diversity [19] was among the most important observations made by these early studies and has been interpreted as evidence for the widespread effect of selection along the genome. This postulate challenged the paradigm of the Nearly Neutral extension of the Neutral Theory [169, 170], which assumes that the large majority of polymorphic and divergent sites are neutral or slightly deleterious. Since then, the search for genes underlying adaptation as well as the quantification of the genome-wide impact of selection has stimulated the development of statistical methods aiming at detecting past adaptive processes from DNA polymorphism data. In 1991, McDonald and Kreitman developed their reference test of selection, and detected adaptation on the *Adh* locus in *Drosophila* [16]. Based on the McDonald and Kreitman test ratios, the fraction α of non-synonymous substitutions driven to fixation by position selection can be estimated by $1-(D_s P_n) / (D_n P_s)$, with $D_s$ and $D_n$ the number of synonymous and non-synonymous substitutions, respectively and $P_s$ and $P_n$ the number of synonymous and non-synonymous polymorphisms, respectively ([171] and see Chapter 6 for more details). Numerous studies have provided evidence for pervasive molecular adaptation in *D. melanogaster*, suggesting that approximately 50% of the amino acid changing substitutions (α=0.5), and similarly large proportions of non-coding substitutions, were adaptive [172–177].

### 4.1. Hitchhiking effects

The first mathematical formulation of the effect of a positively selected allele on intra-specific genetic diversity was proposed by Maynard Smith and Haigh in 1974 and coined the "**hitchhiking model**" [178]. Selection reduces diversity not only at selected sites, but also at linked neutral sites, and the number of variants linked together around a single selected target is inversely proportional to the recombination rate. The hitchhiking model summarizes the relation between the strength of selection on a single adaptive mutant allele, the local recombination rate, and the distribution of surrounding neutral alleles across sites and samples. Under such a linkage model, when a beneficial allele establishes itself in the population, the high rate at which this establishment occurs creates an irregularity in the distribution of neutral alleles around the selected allele. This characteristic signature resulting from positive selection has been coined "**selective sweep**" (hard sweep), a terminology used to describe both the adaptive process and the resulting signal in genetic data. This model served as basis for the development of statistical tools designed to capture the signal of a selective sweep in the presence of different confounding factors ([179, 180] and see Chapter 5 for more discussion on sweep detection). *D. melanogaster* has been among the first organisms for which this approach has been used to map selective sweeps [67, 181, 182], eventually yielding to the identification of several candidate genes/regions for adaptations (Table1) that allowed *D. melanogaster* to extend its geographic range to very heterogeneous environments and to recent anthropogenic changes [183–185]. However, the particular demographic history of the species, and especially the severe founding events followed by population expansion should be considered as a confounding factor, strongly increasing the rate of false positives and thus reducing the performance of sweep detection methods in this specific biological system [186–189]. These insights into the confounding effects of adaptive and neutral processes motivated two lines of research: A) characterizing neutral models accounting for the major demographic events having affected the genome-wide distribution of neutral alleles (see paragraph above) and B) more general formulation of the adaptive process initially described by [178].

**Soft sweep** theory extended the Maynard Smith's and Haigh's hitchhiking model, by including the possibility of (i) recurrent mutations leading to beneficial alleles and (ii) segregating neutral alleles becoming positively selected (i.e. selection from standing variation; reviewed in [190]). Both cases predict an association of the beneficial alleles with several background haplotypes (versus a single one in the hard sweep model). Garud et al. [191] scanned the DGRP dataset to capture signature of hard and soft sweeps, and found a significantly higher number of candidate genomic regions than expected under the neutral admixture model previously calibrated for this population [75]. Furthermore, they found that among their top 50 candidates most cases were better explained by soft than hard sweeps, suggesting that standing genetic variation and recurrence of beneficial alleles play an important role in real-life adaptive processes in *D. melanogaster*. However, the statistical significance of their results is highly dependent on an appropriate calibration of neutral demographic models, suggesting that the performance of soft-sweep detection methods still needs to be tested under a large range of demographic models. In the meantime, the results of genome-wide soft-sweep detection studies should be evaluated carefully when used to support claims about adaptive processes [192].

**4.2. Recurrent hitchhiking and background selection**

Beyond the study of single instances of selective sweeps, *D. melanogaster* and *D. simulans* have also been used to investigate the genome wide effect of recurring sweeps on genetic variation. The relevant model is the recurrent hitchhiking model [193], which describes genome-wide patterns of variation as a function of the occurrence rate of selective sweeps and the distribution of fitness effects of advantageous mutations. Several studies have developed model-based inference approaches to estimate these two parameters using polymorphism and divergence data, reviewed in [194, 195]. All consistent with a strong impact of selection on the pattern of diversity in this species, a wide range of the strength of selection on beneficial mutations ($N_e s$, where $N_e$ is the effective population size and $s$ the selection coefficient) was estimated, ranging from 1–10 [196, 197], ~12 [198], ~40 [199], 350–3500 [67, 171, 200] to ~10,000 [201]. These studies showed that the rate and strength of positive selection was large enough such that a significant amount of neutral alleles in the genome cannot be seen as evolving independently from adaptive sweeping alleles (the dependencies being caused by genetic linkage between beneficial and neutral alleles). Essentially, the disparate estimates reflect variation in the calibration of the different models, in particular according to (i) the type of selection assumed, (ii) the modelled relationship between diversity estimates and selection (strength and frequency) through the action of recombination. These results also revealed the difficulty of telling apart whether genome-wide selection is characterized by a small number of large effect or a large number of small effect adaptive alleles.

The relative importance of positive selection in *Drosophila* has been challenged, however, by studies describing the effect of strictly deleterious alleles on linked neutral variants [68, 202, 203]. This hitchhiking effect caused by selection against recurrent deleterious mutations called **background selection** has been shown to be a valid alternative explanation for low variability in genome regions with low recombination rates [68, 204]. Importantly, Comeron generated a map describing the strength of background selection along the genome as a function of the local recombination and deleterious mutation rate [205]. This study showed that a large proportion (70%) of the observed variation in the level of diversity across autosomes can be explained by background selection alone and therefore called for the inclusion of background selection in further population genomics analyses. Elyashiv et al. recently proposed a method to jointly estimate the parameters of distinct modes of linked selection, accounting for both positive (selective sweeps) and negative background selection [206]. Applied on *D. melanogaster*, they showed that negative selection at linked sites has had an even more drastic effect on diversity patterns in *D. melanogaster* than previously appreciated based on classical selective sweeps models (1.6–2.5-fold). Their results further suggest that 4% of substitutions between *D. melanogaster* and *D. simulans* have experienced strong positive selection ($s \approx 10^{-3.5}$) and that 35% to 45% of substitutions have been weakly selected ($s$ between $10^{-5.5}$ and $10^{-6}$).

### 4.3. Selection on non-coding DNA

Since the 2000s, whole genome comparative analyses accumulated evidence that only a small portion of conserved sequences across species (i.e. potentially functional) was composed of protein-coding genes [207, 208]. In the meantime, genomic surveys identified non-coding genomic sequences showing exceptionally high levels of similarity across species, which were termed **conserved non-coding elements** or CNEs (reviewed in [209]). In *Drosophila*, CNEs are estimated to cover ~30-40% of the genome [207]; Berr et al. 2018). The high levels of evolutionary conservation observed in these regions are postulated to be the result of functional constraints since many CNEs partially overlap with *cis*-regulatory elements [210] and functional non-coding RNAs [211, 212]. In *Drosophila*, several classes of non-coding DNA evolve considerably slower than synonymous sites, and yet show an excess of between-species divergence relative to polymorphism when compared with synonymous sites [174]. While the former observation indicates selective constraints, the latter is a signature of adaptive evolution, which resembles patterns of protein evolution in *Drosophila* [172, 173]. To quantify the intensity and the relative importance of selection in shaping the evolution of non-coding DNA, several studies applied extensions of the McDonald–Kreitman approach, combining polymorphism and divergence analyses. When analysing non-coding DNA in a population from Zimbabwe, Andolfatto estimated that ~20% of nucleotide divergence in introns and intergenic DNA and ~60% in UTRs were driven to fixation by positive selection [174]. Using a hierarchical Bayesian framework, he estimated that significant positive selection acted on non-coding sequences, especially in UTRs [174]. This was recently supported by a whole genome survey of 50 European populations that showed that UTRs and non-coding RNAs are the non-coding genomic regions most subjected to adaptive selection, with >40% of divergence being driven by positive selection (Berr et al.). Specifically focusing on CNEs of the *X* chromosome, Casillas et al. [213] observed a large excess of low-frequency derived SNP alleles within CNE relative to non-CNE regions in an African and two European populations. While low level of purifying and positive selection also act outside of CNEs, Casillas et al. [213] estimated that 85% of the CNEs were functional and evolved under moderately strong purifying selection (*Nes* ~10-100). Altogether, these studies strongly suggest that CNEs are not solely neutral genomic regions with extremely low mutation rates known as mutation "cold spots" [214] but shaped by both purifying selection and adaptive evolution in *Drosophila*. Moreover, these findings support the important role of non-coding regulatory changes in evolution.

**4.4. Selection on synonymous codon usage**

The McDonald & Kreitman test and its extensions are built around the hypothesis that synonymous or four-fold degenerate sites (Note 4) mostly evolve neutrally, while non-synonymous sites are under strong purifying or positive selection. However, both synonymous and four-fold degenerate sites might be subject to selection on synonymous codon usage (see original reference for *Drosophila* by [215], and more recent review by [216]). Comparison of polymorphism and divergence patterns suggested that both strong ($4N_es \gg 1$) and weak ($4N_es \sim 1$) selection applies to synonymous sites in *D. melanogaster* [217, 218]. In this species, the level of codon bias is positively correlated to the levels of expression [219], but negatively correlated to the levels of divergence [220, 221]. Both findings suggest selection on codon usage bias. As in most *Drosophila* species, all preferred codons are GC-ending [219, 222]; selection on codon bias is therefore expected to increase GC content at synonymous sites. Several attempts to detect selection on codon usage bias in *D. melanogaster* have come to conflicting conclusions. Some studies detected evidence for selection favouring GC-ending codons [117, 223], although the intensity of selection may be weaker in *D. melanogaster* compared to other *Drosophila* species [224]. Other studies did not find support for such on-going selection [225, 226], but rather revealed an excess of substitutions toward AT-ending codons. This may either reflect a reduction in selection efficacy ($4N_es$) or a shift in the mutational bias in *D. melanogaster* lineage [227]. The population genetics of codon usage bias can however be affected by confounding, non-adaptive processes such as GC-biased gene conversion ([111] but see section 3.3). In a recent study, Jackson et al. [228] modelled base composition evolution, and found evidence for selection on four fold-degenerate sites along both *D. melanogaster* and *D. simulans* lineages over a substantial period. They showed that while selection intensity on codon usage was rather stable in *D. simulans* in the recent past, it was declining in *D. melanogaster*. In conclusion, the observed AT-biased substitution pattern could not only result from a mutational bias, but likely partially reflects an ancestral reduction in selection intensity.

**4.5. Adaptive chromosomal inversions**

There is ample evidence that inversions play a pivotal role during adaptive processes and various hypotheses have been developed to explain their evolutionary impact [134–136, 229]: (1) According to the "coadaptation" model, inversions have higher fitness and spread because they suppress maladaptive crossing-over which would unlink co-adapted alleles at epistatically interacting loci with high marginal fitness [9, 230]. Genomic analyses in *D. pseudoobscura* support this model and provide evidence that loci in tight linkage with an inversion show epistatic interactions [231]. (2) Under the "local adaptation model", an inversion bears higher fitness because it captures and protects locally adapted loci from recombination with maladaptive migrant haplotypes as initially proposed by [232] and recently revised by [233]. A remarkable conclusion of this model is that the selective advantage of an inversion is determined only by the migration rate of maladapted haplotypes and the amount of linkage among the locally adapted loci. (3) The frequent occurrence of fixed inversions in different species of the genus *Drosophila* [234–237] and in other species groups [238, 239] suggests that many divergent inversions evolved by underdominance and are important components of the speciation process by suppressing gene flow among young sym- or parapatric species [240]. Similarly, inversions play a key role in the evolution of sex chromosomes by keeping together alleles in sex determining factors and sexually antagonistic genes [241]. (4) Conversely, inversions can also be maintained due to overdominance or other types of balancing selection. In line with this model, many inversions, particularly in *Drosophila*, are commonly found to segregate at intermediate frequencies in natural and experimental populations [242].

**4.6. Adaptive insertions of tranposons**

Like other type of mutations, TE insertions are expected to be mostly deleterious or evolutionary neutral. However, some transposable elements could be beneficial and positively selected. There are several possible mechanisms by which a TE can be advantageous; either by directly affecting the gene function of individual genes, or by modifying regulatory elements [243, 244]. Due to recent technical advances in sequencing technology (NGS) and due to the rapidly growing number of whole genome data, the ability to detect selected TE insertions has considerably increased in the past few years. Different methods have been developed to infer selection acting on TE insertions. Villanueva-Cañas et al. [245] provide a detailed overview over the main approaches and their specificities: (1) **DNA sequence conservation** analyses can be used to detect past events of domestication of TEs as regulatory elements, where TE insertions are conserved among closely related species due to purifying selection (see for example Berr et al). (2) Methods developed to detect **selection on linked polymorphisms** from SNPs (see 4.1 section and Chapter 5 for more discussion) can also be applied to identify positively selected TEs. Based on either a bias in frequency spectra or haplotype structure, over 35 putatively adaptive TEs were identified in genome-wide studies in *D. melanogaster* to date [160, 164, 246, 247]. (3) A third method is built around **environmental association analyses** that include genome scans for selection performed in parallel in populations from different environments to detect specific adaptation driven by environmental conditions. Using this approach, González et al. [248] discovered several recent TE insertions in *D. melanogaster* that are putatively involved in local adaptation. These TEs exhibit low population frequencies in ancestral population (Africa) but are common in derived populations (North America and Australia). (4) Using a coalescent framework approach, Blumenstiel et al. [168] identified seven additional putative adaptive insertions exhibiting higher population frequency than expected according to their **estimated allele age**. (5) Finally, selection on TE insertions should be validated at the phenotypic level using **functional assays** to identify the molecular and fitness effects. One well-documented example is the insertion *Bari-Jheh* that was found to affect the level of expression of its nearby genes under oxidative stress conditions and to increase resistance to this stress [249, 250].

Beside the impact of single TE insertions, there is growing evidence for a more global effect of TEs on molecular functions. Especially, in *Drosophila*, TEs seem to play a role in a diversity of cellular processes [160], such as the establishment of dosage compensation [251], heterochromatin assembly [252] and brain genomic heterogeneity [253].

### 4.7. Faster-X evolution

According to a theory proposed by Charlesworth et al. [254], the rates of evolution of *X*-linked loci are expected to be faster than autosomal ones if mutations are partially recessive ($0<h<1/2$, with $h$ the coefficient of dominance) and expressed in both sexes or males only. In heterogametic males (*XY*), *X*-linked mutations are hemizygous and therefore directly exposed to selection, whereas new recessive autosomal mutations are masked from expression in heterozygotes individuals. Moreover, the effective recombination rate is ~1.8-fold greater on the *X* compared to autosomes [194], which reduces Hill-Robertson interference and increases the efficiency of selection. The increased selection in hemizygous males together with the higher efficiency of selection due to the increased recombination may act synergistically to account for the "faster-*X* evolution", which is generally supported by genomic data collected in *Drosophila* populations (reviewed in [255]). Levels of polymorphism are similar on *X*-linked loci to autosomal ones in African populations, but lower in derived populations [30, 50], which might be a consequence of selective sweeps in response to the adaptation of new environments [256]. However, recombination seems to play a secondary role in determining pattern of diversity along the *X*-chromosome. Contrary to autosomes, the *X*- chromosome exhibits global nucleotide diversity only weakly correlated with recombination rate (Figure 2), and a non-synonymous diversity completely independent [257].

In contrast with polymorphism, divergence among *Drosophila* relatives is greater for the *X* than for autosomes (reviewed in [255]). Higher efficiency of selection on *X* is supported by the estimated higher percentage of sites undergoing both strongly deleterious and adaptive evolution than autosomes, and a lower level of weak negative selection in *D. melanogaster* [45, 46, 257]. Codon usage bias in *Drosophila* is also higher for *X*-linked genes than for autosomal ones, possibly due to the higher effective recombination rate and their resulting reduced susceptibility to Hill–Robertson effects [117]. In the end, faster-*X* evolution also implies that genes for reproductive isolation have a higher probability of being *X*-linked, what is generally true [160].

## 5. Perspectives: temporal and geographical clines

Organisms with broad geographic distributions, such as various species of the genus *Drosophila*, are commonly found along environmental gradients. Such transects have long been in the focus of evolutionary geneticists, as they provide natural test beds to investigate the evolutionary underpinnings of local adaptation [258]. Studying spatially or temporally changing genotypes and phenotypes, which are commonly referred to as "clines" [259], has a long history in *D. melanogaster* [260–262]. While there is growing evidence for longitudinal clines in Africa [263] and in Europe [43], most data have been collected from latitudinal gradients along the North American and Australian east coasts. A large body of literature documents steep and persistent clines in many fitness-related phenotypes, which are often recapitulated on multiple continents. These include, for example, clines in body-, wing- and organ-size [264–268], lifetime fecundity and lifespan [269] as well as heat and cold resistance [270–272]. Similarly, various genetic polymorphisms such as microsatellites [273], SNPs [274], TEs [246] and inversions [140–142] have been found to vary clinally. Besides these well-defined spatial clines there is growing evidence for rapid adaptation on seasonal timescales leading to temporal clines. These are characterized by predictable annual fluctuations in allele frequencies [20] and variation in life history traits [21] and innate immunity [275].

Ongoing advances in next-generation sequencing technology prompted the development of analytical methods to identify putative targets of local and clinal adaptation (reviewed in [276]) and only recently allowed to extend the hunt for clinal genetic variation from single loci to genome-wide scales. A rapidly increasing number of studies in *D. melanogaster* have started to comprehensively investigate clinal genomic patterns - mostly by comparing the endpoints of latitudinal gradients from the Australian and North American east coasts [58–60, 274]. Many of these pioneering studies identified common patterns in the distribution of genome-wide clinal variation which provide insights, but also raise new questions about the evolutionary mechanisms involved in adaptation: (1) Loci with extensive clinal differentiation are not homogeneously distributed along the genome, but strongly clustered within large inversions [58, 60, 140], which suggests that inversions play an important role during local adaptation – potentially by keeping together co-adapted loci associated with polygenic trait variation [232, 233]. However, the identity of these loci and the affected traits remain largely unknown so far. (2) Many clinal polymorphisms, such as the chromosomal inversion *In(3R)Payne* and variants of the alcohol dehydrogenase (*Adh*) locus, are paralleled on multiple continents [33, 58, 59, 274] and change frequencies in a predictable fashion. While parallel adaptive evolution due to spatially varying selection along analogous environmental gradients on different continents may shape many clinal patterns, other non-adaptive evolutionary forces could have similar effects. For example, a handful of studies found independent evidence for varying levels of admixture with African genetic variation both in North America [54, 75, 76] and Australia [77]. These findings highlight that clines, which are often considered to be the prime outcome of spatially or temporally varying selection, are potentially confounded with neutral evolutionary processes such as spatially restricted gene flow or admixture [277]. At last, (3) all aforementioned studies failed to identify large numbers of clinal loci with large or even fixed allele differences at the opposite endpoints of the latitudinal gradients. For example, no more than 0.1% of all SNPs exhibited allele frequency differences > 0.5 between Florida and Maine, while not a single SNP exceed an allele frequency difference of 0.92 in the analyses of [58]. These findings are consistent with observations from other *Drosophila* species, which also found moderate and gradual clinal allele frequency changes [139, 278, 279], but in stark contrast to common model expectations for clinal evolution [280]. Together, these first analyses of clinal genomic data clearly show that it still remains challenging to disentangle the evolutionary contribution of selection and demography to clinal variation in natural populations.

Efforts of two large population genomic consortia are currently underway to densely sample natural populations through time and space both in North America [57] and Europe [43]. These comprehensive datasets will markedly extend earlier efforts that focused mostly on the comparison of clinal endpoints. Particularly the analyses of previously largely ignored European *D. melanogaster* populations will allow to make clear predictions about the adaptive process in derived populations from North America and Australia.

## Acknowledgements

We are very grateful to Laurent Duret, Jeffrey Jensen, Gabriel Marais, Alan Moses, Cristina Vieira and Stephen Wright for their helpful comments on the manuscript. We are very thankful to Julien Dutheil for the invitation to write this chapter.

## Notes

**Note 1**: the mathematical framework that integrated Darwin's theory of evolution and the mechanisms of heredity discovered by Mendel.

**Note 2**: defined as genome scale analyses of polymorphism including polymophism/divergence comparisons, but not analyses strictly based on divergence.

**Note 3**: Ectopic recombination: recombination between two similar nonhomologous sequences, i.e. two TE copies of the same family inserted at different genomic locations. Such DNA exchange between non-orthologous regions leads to chromosomal rearrangement.

**Note 4**: four-fold degenerate sites consist in sites for which all four possible nucleotides at this position would encode for the same amino acid, representing a subset of all synonymous sites

**Figure 1: Map illustrating world-wide distribution, migration routes and clinal differentiation of the cosmopolitan species *D. melanogaster*.** Populations are separated in ancestral (red), ancient (orange) and newly-introduced (blue) populations, according to the categorization in David and Capy [4]. The expected ancestral range (Zambia) is highlighted in dark red. Primary colonization routes across populations are shown by colored arrows: the European colonization started

approximately ~10-19,000 years ago [73, 75], followed by a spread to Asia ~5,000 years ago [73] and a more recent range expansion to Australia and North America within the last 200 years [2]. Patterns of recent admixture (dotted grey arrows) were documented from European alleles in Africa [50], and from African alleles to North America and Australia [77, 78]. Clinal genetic and phenotypic differentiation (dash-dotted black arrows and color gradient) are documented along latitudinal gradients in North America [58, 77] and Australia [60, 77], and along longitudinal gradients in Africa [263] and in Europe [43]. At the time of the review, no information about demography of South-American populations was available. The dark grey areas depict expected habitable geographic regions and were modelled from 4951 unique world-wide sampling-points in the TaxoDros database (http://taxodros.uzh.ch) and climatic data from the WorldClim database (http://worldclim.org) using the *R*-package *dismo* (http://rspatial.org/sdm/). Note that distribution models can be confounded by unequal sampling and may thus explain the missing predicted distribution in South-Western Africa, Central Asia and Russia.

**Figure 2: Correlation between nucleotide diversity and recombination**. Nucleotide diversity ($\pi$) calculated in 10kb non-overlapping windows was estimated for 48 European populations (DrosEU data, [43] and compared to recombination rate (*r*) obtained from [93] for the four autosomal arms (*2L*, *2R*, *3L* and *3R*) and the *X* chromosome. We therefore averaged $\pi$ in equally-sized bins according to discrete log-transformed values of *r* observed in the corresponding genomic regions. The shaded polygons surrounding average values (central lines) for each of the 48 populations show the 95% confidence intervals.

**Table 1. Documented selective sweeps in African and non African populations of *D. melanogaster***

| Gene(s) involved in the sweep | Sweep size (kb) | Populations | Biological Function | Reference |
|---|---|---|---|---|
| *Acetylcholineesterase* (*Ace*) | ~1.5 | Non-African populations | Insecticide resistance | Karasov et al. 2010; Messer and Petrov 2013; Kapun et al. 2018 |
| *Argonaute-2* (*AGO2*) | >50 | *D. melanogaster*, *D. simulans* and *D. yakuba* | Resistance to viral infection | Obbard et al. 2011 |
| *brinker* gene (*brk*) | 83-124 | European population | Cold tolerance | Glinka et al. 2006; Wilches et al. 2014 |
| *CG18 508* and *Fcp3C* | 14 | Non-African populations | | DuMont and Aquadro 2005 |
| *CHKov1* | ~25 | Non-African populations | Resistance to viral infection | Magwire et al. 2011 |
| *Cyp6g1* | | Non-African populations | Insecticide resistance | Schmidt et al. 2010; Battlay et al. 2016, 2018; Kapun et al. 2018 |
| *diminutive (dm)* | 25 | African and non-African populations | Positive regulator of body size | Jensen et al. 2007 |

| Gene | | | | |
|---|---|---|---|---|
| *Fezzik (fiz)* | 1.8 | European population | Growth | Saminadin-Peter et al. 2012; Glaser-Schmitt and Parsch 2018 |
| *HDAC6* | 2.7 | African population | Stress surveillance | Svetec et al. 2009 |
| *Notch* | 14 | Non-African populations | Development | DuMont and Aquadro 2005 |
| *phantom* (*phm*) | 12-20 | European population | Cytochrome P450 enzyme | Orengo and Aguade 2007 |
| *polyhomeotic-proximal* (*ph-p*) | 30 | European population | Reduced temperature-induced plasticity | Beisswanger and Stephan 2008; Voigt et al. 2015 |
| *roughest* (*rst*) | 0.361 | African population | Apoptosis | Pool et al. 2006 |
| *Suppressor of Hairless* (*Su[H]*) | 1.2 | African population | Growth; *Notch* signalling | Depaulis et al. 1999 |
| *wings apart-like* (*wapl*) | ~60 | European populations | Chromatin organization | Beisswanger et al. 2006 |
| > 50 candidates | | North-American population | | Garud et al. 2015 |

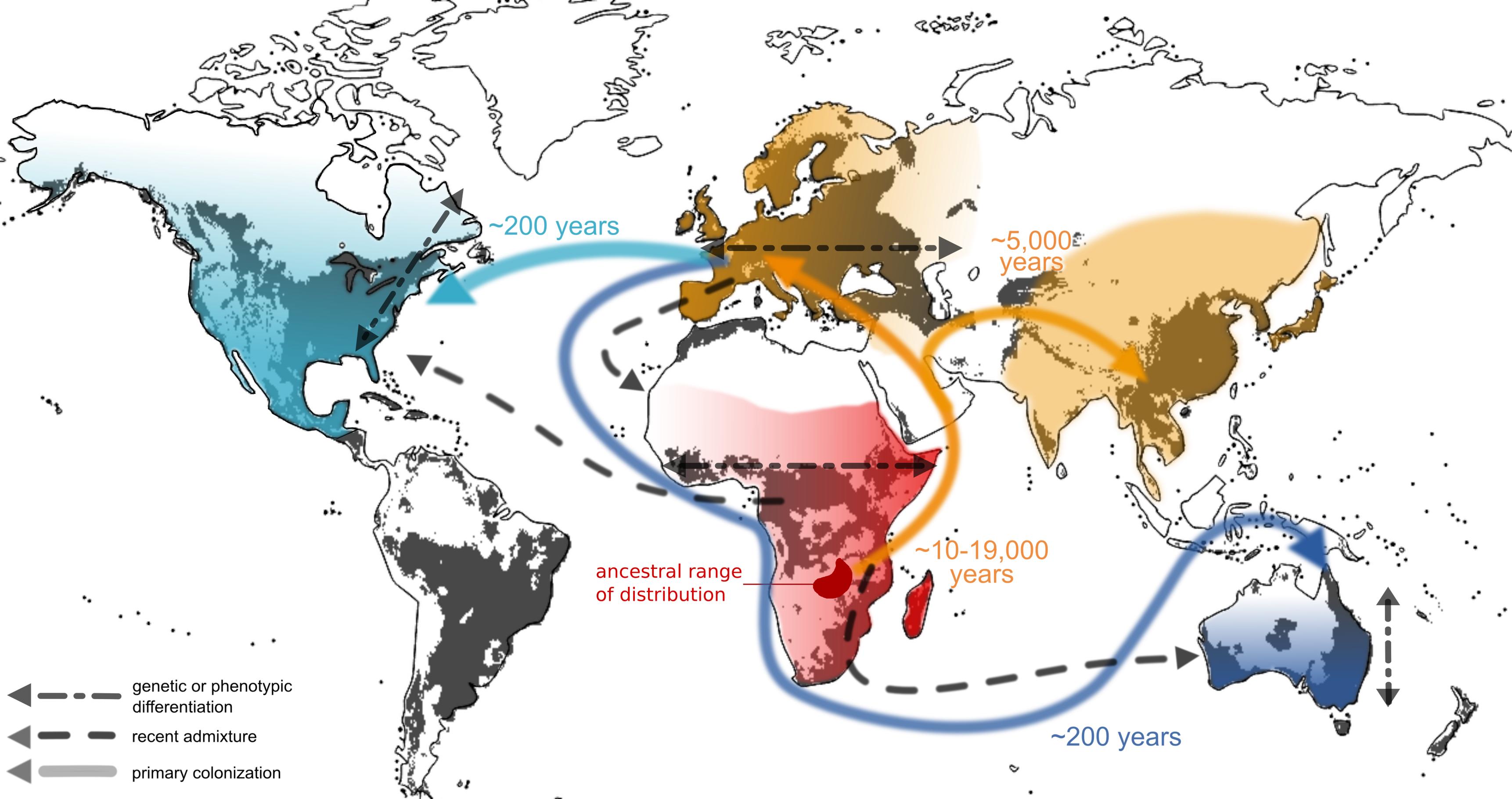

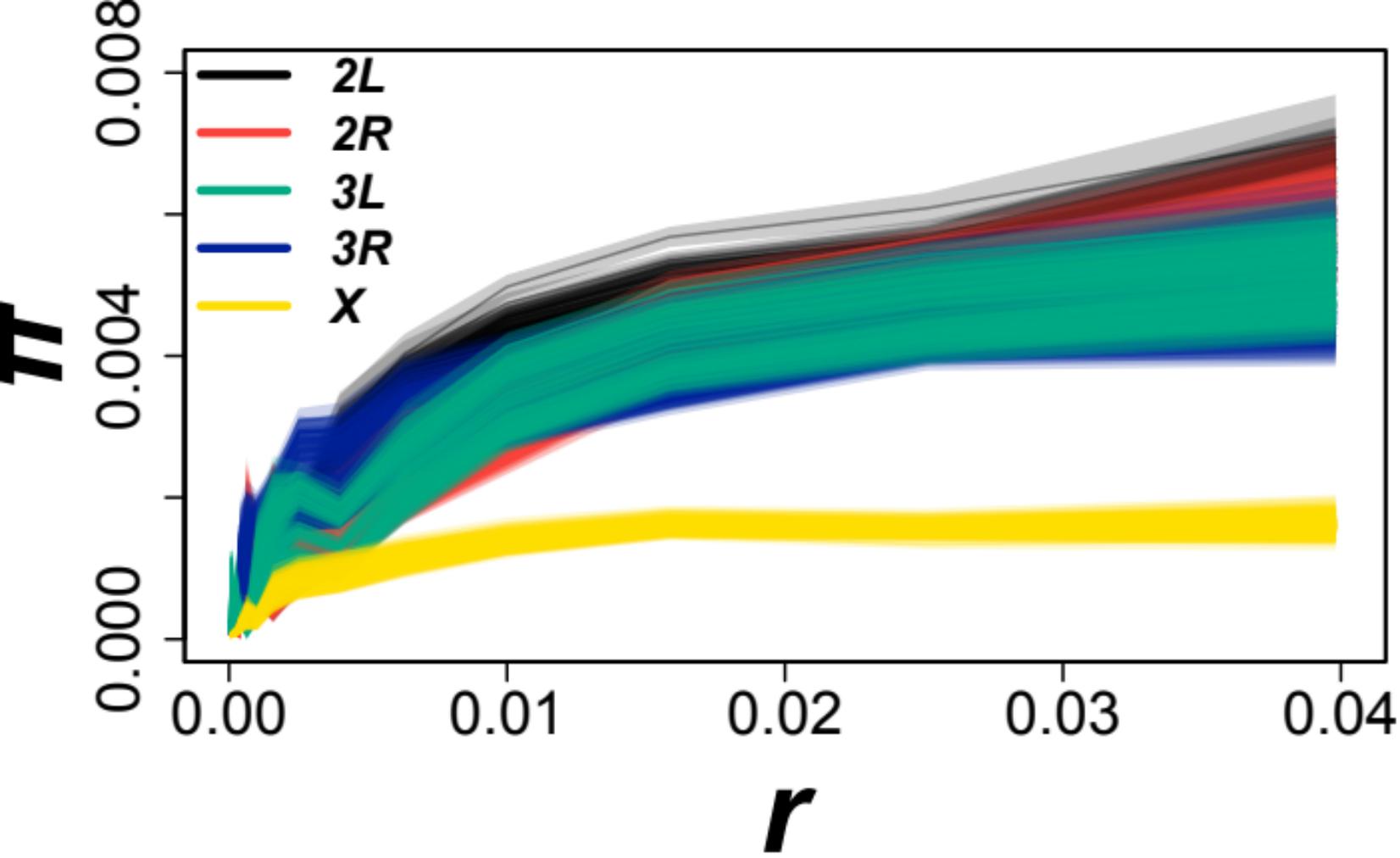